# Electronic Transport in a Topological Semimetal $WTe_2$ Single Crystal

A. N. Perevalova[1], S. V. Naumov[1], V. V. Chistyakov[1], E. B. Marchenkova[1], B. M. Fominykh[1,2], and V. V. Marchenkov[1,2,*]

[1]*Mikheev Institute of Metal Physics, Ural Branch, Russian Academy of Sciences, Ekaterinburg, 620108 Russia*
[2]*Ural Federal University, Ekaterinburg, 620002 Russia*

*e-mail: march@imp.uran.ru

**Abstract**

Electrical resistivity, magnetoresistivity, and the Hall effect have been studied in a topological semimetal $WTe_2$ single crystal in the temperature range from 12 to 200 K under magnetic fields up to 9 T. It has been found that quadratic temperature dependences of the electrical resistivity in the absence of a magnetic field and the conductivity in a magnetic field are observed at low temperatures, which is apparently associated with contributions from various scattering mechanisms. Single-band and two-band models were used to analyze data on the Hall effect and magnetoresistivity. These results indicate electron–hole compensation with a slight predominance of electron charge carriers.

## 1. Introduction

In recent years, a large number of different topological materials have been discovered, including topological insulators and topological semimetals [1–6]. Considerable attention is drawn to topological semimetals that can be divided into the following three main groups: Weyl semimetals, Dirac semimetals, and topological nodal line semimetals. In Dirac and Weyl semimetals, two double-degenerate bands or two nondegenerate bands intersect with each other at special points (nodes) near the Fermi level, thereby forming Dirac points or Weyl points, respectively, and disperse linearly in all three directions of momentum. The corresponding low-energy excitations behave similarly to Dirac and Weyl fermions in high-energy physics.

It is known that Weyl fermions can occur in systems with broken inversion or time-reversal symmetry [3–6]. In this regard, noncentrosymmetric semimetals, such as TaAs, and magnetic semimetals, such as some Heusler alloys, are candidates for Weyl semimetals. The first experimental confirmation of the existence of the Weyl semimetal phase was obtained on single crystals of the TaAs family (TaAs, TaP, NbAs, and NbP) in 2015 [7]. Moreover, the authors of [8] predicted a special type of band crossing with a strongly tilted Weyl cone along a certain direction in the momentum space, the so-called type-II Weyl semimetals. The existence of the type-II Weyl semimetal phase was predicted and experimentally confirmed in such layered transition metal dichalcogenides as $WTe_2$ [8, 9], $MoTe_2$, and ternary compound $Mo_xW_{1-x}Te_2$ [10].

The peculiarities of the electronic structure of topological materials are reflected in their electronic properties and lead to a number of unusual effects, such as extremely large magnetoresistance without a tendency to saturation, high mobility and low effective mass of current carriers, nontrivial Berry phase, chiral anomaly, anomalous Hall effect, and linear behavior of optical conductivity [5, 6]. One of these unusual effects is a quadratic temperature dependence of the electrical resistivity of $WTe_2$ [11] and $MoTe_2$ [12] in a very wide temperature range from 2 to 70 K and 50 K, respectively. It can be expected that a quadratic temperature dependence of their resistivity should be observed in the presence of a magnetic field as well. It is also worth noting that either single-band [13] or two-band models are usually used when analyzing data on the Hall effect with subsequent calculation of the concentration and mobility of current carriers [14]. At the same time, it is not entirely clear how correct a particular model is.

This study is devoted to the analysis of kinetic properties (electrical resistivity, magnetoresistivity, the Hall effect) of a WTe$_2$ single crystal to establish the form of temperature dependence of resistivity (conductivity) in a magnetic field, and to the use of single-band and two-band models for the analysis of galvanomagnetic properties.

## 2. Materials and techniques

### 2.1. Single crystal growth and structural analysis

WTe$_2$ single crystals were grown using the method of chemical vapor transport [15]. Fig. 1 shows the scheme of growing the single crystals. Tungsten and tellurium in a stoichiometric ratio were placed in a quartz ampoule with a length of 24 cm and a diameter of 1.5 cm. The ampoule was evacuated to a residual pressure of about $10^{-4}$ atm and then placed in a horizontal tube furnace with a linear temperature gradient. The hot zone had a temperature of 850°C, and the cold zone, i.e., the crystal growth zone, had a temperature of 770°C. The single crystal growth process lasted 500 h. The resulting crystals have a needle shape with lengths of 3 – 5 mm, widths of 0.2 – 1.0 mm, and thicknesses of 50 – 400 μm.

Fig. 2 shows a fragment of the diffraction pattern taken from the surface of the WTe$_2$ sample. All peaks can be indexed as (00$l$). Hence, the surface of a WTe$_2$ single crystal coincides with the (001) plane. It was established that WTe$_2$ compound crystallized in the orthorhombic structure (space group $Pmn2_1$) with lattice parameters $a = 3.435(8)$ Å, $b = 6.312(7)$ Å, and $c = 14.070(4)$ Å.

The surface microstructure and the chemical composition of the crystals were analyzed on a FEI Quanta 200 Pegasus scanning electron microscope equipped with an EDAX attachment for X-ray energy dispersive microanalysis at the Collaborative Access Center "Testing Center of Nanotechnology and Advanced Materials" (TC NTAM), Institute of Metal Physics, Ural Branch, Russian Academy of Sciences. Fig. 3 shows images of the (001) surface and the lateral surface the WTe$_2$ single crystal. As can be seen from Fig. 3, the resulting single crystal has a layered structure. Fig. 4 shows the results of X-ray energy dispersive microanalysis of the WTe$_2$ single crystal. The contents of W and Te in the compound are 33.17 and 66.83 at %, respectively. Thus, the chemical composition of the single crystal corresponds to the stoichiometric ratio WTe$_2$.

### 2.2. Techniques for measuring kinetic properties

The resistivity and the Hall effect were measured using the four-contact method in the temperature range from 12 to 200 K and in magnetic fields up to 9 T on a PPMS-9 universal system for measuring physical properties (Quantum Design, USA) at the TC NTAM of the Institute of Metal Physics, Ural Branch, Russian Academy of Sciences. The electrical contacts were prepared using thin copper wire and silver paste. Measurements were performed with an electric current that flowed in the (001) plane and the magnetic field direction perpendicular to this plane. The ratio of resistivities of the WTe$_2$ single crystal under study at room and helium temperatures is $\rho_{300\,K}/\rho_{4.2\,K} \approx 55$, which indicates its high electrical purity.

In this study, the electrical resistivity in the absence of a magnetic field is denoted by $\rho$ or $\rho(0)$, the magnetoresistivity is denoted as $\Delta\rho_{xx} = \rho(B) - \rho(0)$ (where $\rho(B)$ is the resistivity in magnetic field $B$), and the Hall resistivity is denoted by $\rho_H$. For ease of interpretation and presentation of experimental results, some of them are given in the form of magnetoconductivity $\sigma_{xx}=\Delta\rho_{xx}/(\Delta\rho_{xx}^2+\rho_H^2)$.

## 3. Results and discussion

### 3.1. Electrical resistivity

Fig. 5 shows the temperature dependence of the electrical resistivity $\rho(T)$ of the WTe$_2$ single crystal. In the temperature range from 12 to 70 K, this dependence can be expressed as follows

$$\rho = \rho_0 + AT^2. \quad (1)$$

The quadratic temperature dependence of the electrical resistivity was observed in pure metals [16]. As a rule, the contribution $\sim T^2$ is associated with electron–electron scattering that is usually observed at temperatures below 10–15 K [16, 17]. At higher temperatures, electron–phonon scattering mechanism should prevail, which leads to the dependence $\rho(T) \sim T^5$ at $T \ll \Theta_D$ ($\Theta_D$ is the Debye temperature) and to the linear dependence $\rho(T)$ at temperatures comparable to $\Theta_D$. The Debye temperature for WTe$_2$ is 133.8 ± 0.06 K [18]. In our case, the resistivity contribution proportional to $T^5$ is not observed at low temperatures $T \ll \Theta_D$, i.e., the Bloch–Grüneisen law is not followed.

The quadratic behavior of the dependence $\rho(T)$ at temperatures between 12 K and 70 K can be explained as follows. According to Drude formula, the conductivity can be expressed as

$$\sigma = \frac{ne^2\tau}{m} = \frac{ne^2l}{mv}, \quad (2)$$

where $n$ is the concentration of current carriers; $e$ is the elementary charge; $\tau$ is the relaxation time; $m$ is the electron mass; and $l$ and $v$ are the mean free path and the velocity of conduction electrons, respectively. In [19], the $l$ value was estimated for our crystal, and it was demonstrated that the mean free path is $l = \text{const} + CT^{-2}$ in the temperature region (24−55) K, which is consistent with the quadratic behavior of the dependence of the electrical resistivity at temperatures up to 70 K. It can be assumed that contributions from various scattering mechanisms lead to the quadratic nature of the dependence $\rho(T)$ due to the peculiarities of the electronic structure of WTe$_2$ at $T \leq 70$ K. This should also manifest itself in the resistivity (conductivity) measured in the presence of a magnetic field.

### 3.2. Magnetoresistivity

Fig. 6a shows the field dependence of the magnetoresistivity $\Delta\rho_{xx} = \rho(B) - \rho(0)$ of the WTe$_2$ single crystal at a temperature of $T = 12$ K. As can be seen from Fig. 6a, the magnetoresistivity $\Delta\rho_{xx}$ varies with variation of the field according to a law close to quadratic $\Delta\rho_{xx} \sim B^n$, where $n \approx 1.93 \pm 0.01$. Such behavior is typical for compensated conductors with a closed Fermi surface in the region of high effective magnetic fields ($\omega_c\tau \gg 1$, where $\omega_c$ is the cyclotron frequency) [17].

Fig. 6b shows the temperature dependence of the resistivity $\rho(T)$ of the WTe$_2$ single crystal in a magnetic field of 9 T. There is a minimum in the $\rho(T)$ curve. A similar dependence was observed, for example, in tungsten single crystals [20], in which the presence of a minimum is explained by the transition from high effective magnetic fields to weak ones. According to [17], the conductivity of a compensated metal with a closed Fermi surface in the region of high effective magnetic fields ($\omega_c\tau \gg 1$) is determined by the contributions from various scattering mechanisms. Therefore, it is more convenient to carry out further analysis on the basis of the dependence of conductivity $\sigma_{xx}$ in a magnetic field. To simplify the calculations, the formula of $\sigma_{xx}$ for the case of an isotropic crystal is further used, in which $\sigma_{xx}$ is related to components $\Delta\rho_{xx}$ and $\rho_H$ of the resistivity tensor as $\sigma_{xx}=\Delta\rho_{xx}/(\Delta\rho_{xx}^2+\rho_H^2)$. The dependence of $\sigma_{xx}$ on $T^2$ is shown in the inset of Fig. 6b, from which one can see that conductivity $\sigma_{xx}$ in a magnetic field also changes with temperature according to a quadratic law, but in a narrower temperature range from 12 to about 55 K when compared to electrical resistivity. Thus, quadratic temperature dependences are observed for both the electrical resistivity in the absence of a magnetic field and the conductivity in a magnetic field, which is apparently associated with contributions from various scattering mechanisms.

*3.3. The Hall effect*

Fig. 7a shows the temperature dependences of the Hall coefficient $R_H$, and the concentration $n$ and mobility $\mu$ of the main charge carriers in the WTe$_2$ single crystal that were obtained within of a single-band model by formulas

$$R_H = \frac{\rho_H}{B}, \qquad (3)$$

$$n = \frac{1}{R_H \cdot e}, \qquad (4)$$

$$\mu = \frac{R_H}{\rho}. \qquad (5)$$

Given that $R_H < 0$, the main charge carriers are electrons with the concentration of $n \approx 5.3 \times 10^{19}$ cm$^{-3}$ and the mobility of $\mu \approx 5.9 \times 10^3$ cm$^2$/(V s) at $T = 12$ K. The $n$ value determined by formula (4) changes weakly with temperature, which is typical for a number of compensated conductors with a closed Fermi surface [21, 22]. At the same time, the mobility $\mu$ calculated by formula (5) strongly decreases with temperature, which can be explained by an increase in the scattering efficiency of current carriers.

As was shown in [19], the Hall resistivity $\rho_H$ of WTe$_2$ is nonlinearly dependent on the magnetic field $B$. It is assumed that such behavior of the dependence $\rho_H(B)$ may be related to the mechanism of scattering of conduction electrons on the sample surface. This was observed in [23, 24] where compensated metals with a closed Fermi surface were investigated under conditions of a static skin effect. A strongly nonlinear dependence of $\rho_H$ on the field in WTe$_2$ was also observed at low temperatures in [14, 25], which was explained by the presence of current carriers of the electron and hole types. In systems containing electron and hole charge carriers, a two-band model is typically used to analyze the field dependences of resistivity $\rho$ in a magnetic field and Hall resistivity $\rho_H$. The expressions for $\rho$ and $\rho_H$ are written in the following form given in [14]

$$\rho = \frac{1}{e} \frac{(n_h\mu_h+n_e\mu_e)+(n_h\mu_e+n_e\mu_h)\mu_h\mu_e B^2}{(n_h\mu_h+n_e\mu_e)^2+(n_h-n_e)^2\mu_h^2\mu_e^2 B^2}, \qquad (6)$$

$$\rho_H = \frac{B}{e} \frac{(n_h\mu_h^2-n_e\mu_e^2)+(n_h-n_e)\mu_h^2\mu_e^2 B^2}{(n_h\mu_h+n_e\mu_e)^2+(n_h-n_e)^2\mu_h^2\mu_e^2 B^2}, \qquad (7)$$

where $n_e$ ($\mu_e$) and $n_h$ ($\mu_h$) are the concentration (mobilities) of electrons and holes, respectively. As shown in Fig. 7b, the field dependences of the resistivity $\rho(B)$ in a magnetic field and the Hall resistivity $\rho_H(B)$ for WTe$_2$ at 12 K were described using the two-band model by formulas (6) and (7), respectively. The following values of the concentrations and mobilities of electrons and holes were obtained: $n_e = (3.14 \pm 0.01) \times 10^{19}$ cm$^{-3}$, $n_h = (2.78 \pm 0.01) \times 10^{19}$ cm$^{-3}$, $\mu_e = (4.77 \pm 0.02) \times 10^3$ cm$^2$/(V s), and $\mu_h = (3.42 \pm 0.01) \times 10^3$ cm$^2$/(V s). Proportion $n_e \approx n_h$ points to electron–hole compensation in WTe$_2$.

Thus, estimates of the concentrations and mobilities of current carriers obtained using both single-band and two-band models are in good agreement with each other. This also applies to the values of the Hall coefficient, which equal $R_H = -1.168 \times 10^{-1}$ cm$^3$/C and $R_H = -1.170 \times 10^{-1}$ cm$^3$/C in the case of single-band and two-band models, respectively.

## 4. Conclusions

Studies of the kinetic properties of the topological semimetal WTe$_2$ single crystal have shown that both the electrical resistivity in the absence of a magnetic field and the conductivity in the field depend on a temperature according to the quadratic law in a wide temperature range from 12 K to 70 and 55 K, respectively. This appears to be associated with the contributions from various scattering mechanisms.

As a result of the analysis of experimental data on the Hall effect and the resistivity in a magnetic field, the concentrations and mobilities of current carriers in WTe$_2$ were estimated using both single-band and two-band models. These results are in good correlation and point to electron–hole compensation with a slight predominance of electron charge carriers.

**Acknowledgments**

The authors thank N.G. Bebenin for the useful discussion of the obtained results and valuable advices on their presentation.

Electrical resistivity studies (sect. 3.1) were performed within the framework of State assignment of the Ministry of Science and Higher Education of the Russian Federation (theme Spin, no. 122021000036-3), supported in part by the Scholarship of the President of the Russian Federation to young scientists and graduate students (A.N.P., SP-2705.2022.1). Studies on the magnetoresistivity (sect. 3.2) and the Hall effect (sect. 3.3) were supported by the Russian Science Foundation (grant no. 22-42-02021).

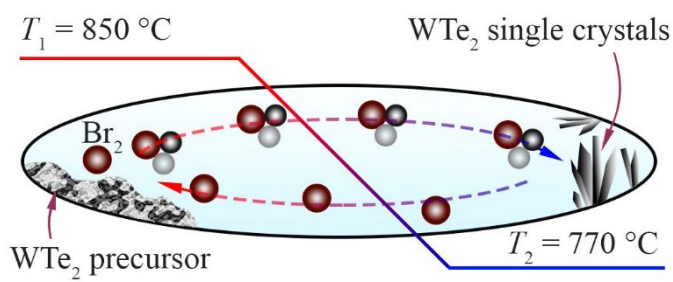

**Figure 1.** The scheme of growing WTe$_2$ single crystals by the chemical vapor transport method with Br$_2$ as a transport agent.

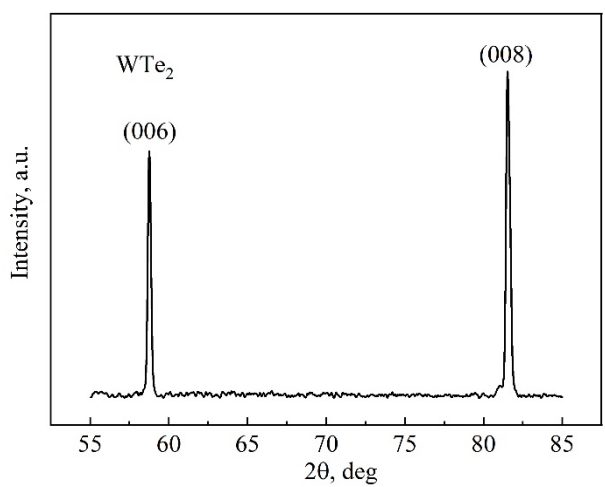

**Figure 2.** A fragment of a diffraction pattern (Cr$K\alpha$) taken from the surface of a WTe$_2$ single crystal.

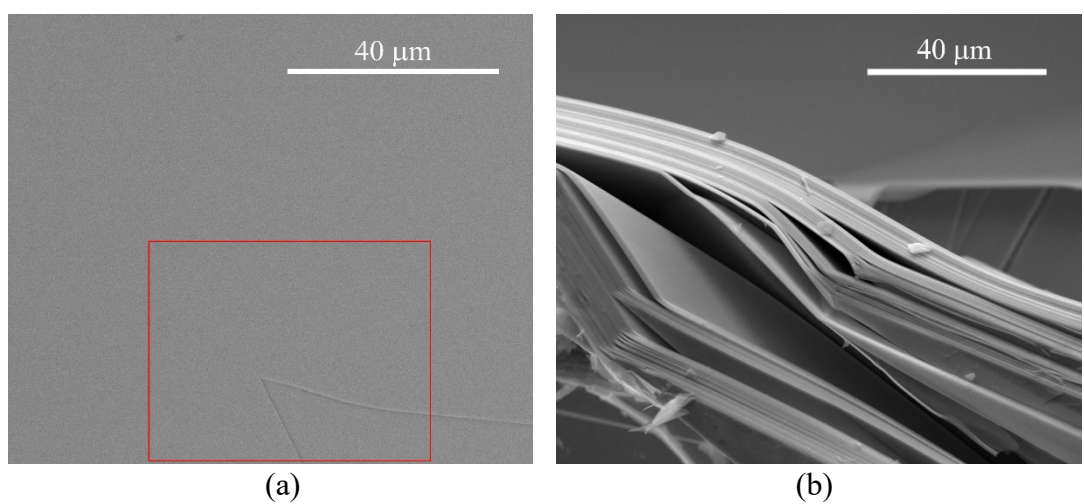

**Figure 3.** The microstructure of the surface of the WTe$_2$ single crystal: (a) the surface of the (001) type, (b) the lateral surface of the sample. The region in which the chemical composition of the sample was investigated is highlighted in Fig. 3a.

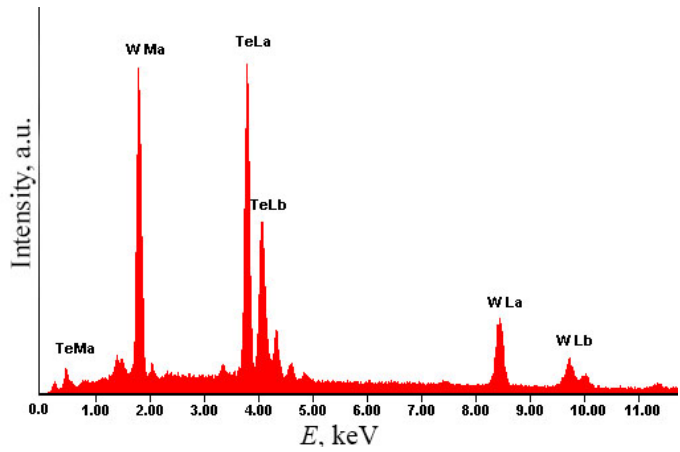

**Figure 4.** Analysis of the chemical composition of the WTe$_2$ single crystal in the region highlighted in Fig. 3a. The contents of W and Te is are 33.17 and 66.83 at %, respectively.

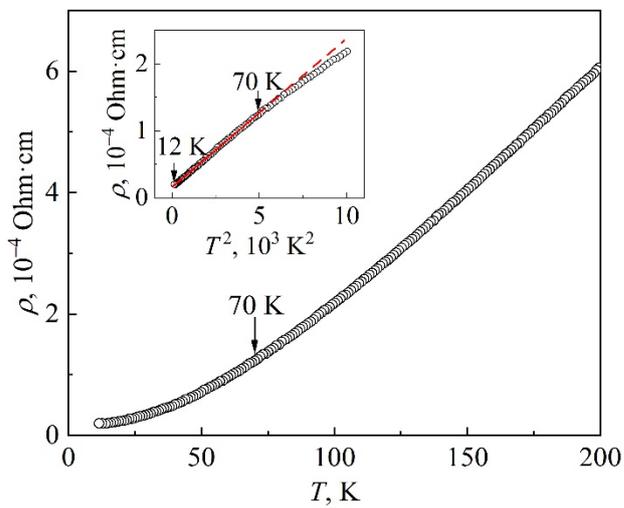

**Figure 5.** The temperature dependence of the electrical resistivity ρ(T) of WTe$_2$ in the temperature range from 12 to 200 K. The inset shows the dependence ρ = f(T$^2$) in the temperatures from 12 to 100 K.

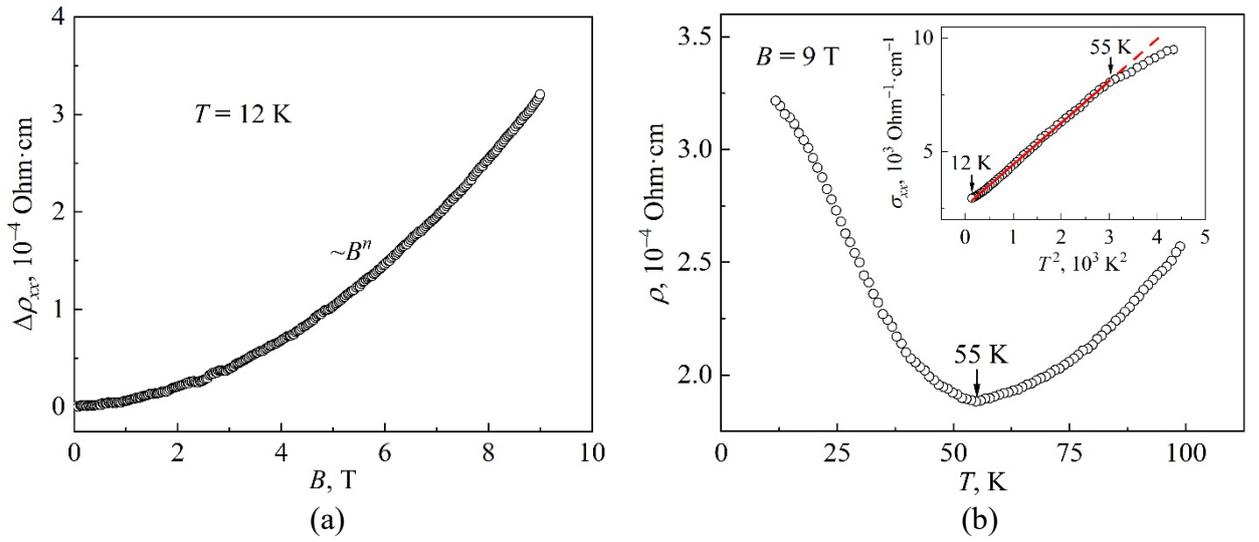

**Figure 6.** (a) The field dependence of the magnetoresistivity $\Delta\rho_{xx}(B)$ of WTe$_2$ at $T = 12$ K. (b) Temperature dependence of the resistivity $\rho(T)$ of WTe$_2$ in a magnetic field of 9 T in the temperature range from 12 to 100 K. The inset shows the conductivity $\sigma_{xx} = f(T^2)$ in a magnetic field of 9 T in the temperature range from 12 to 65 K.

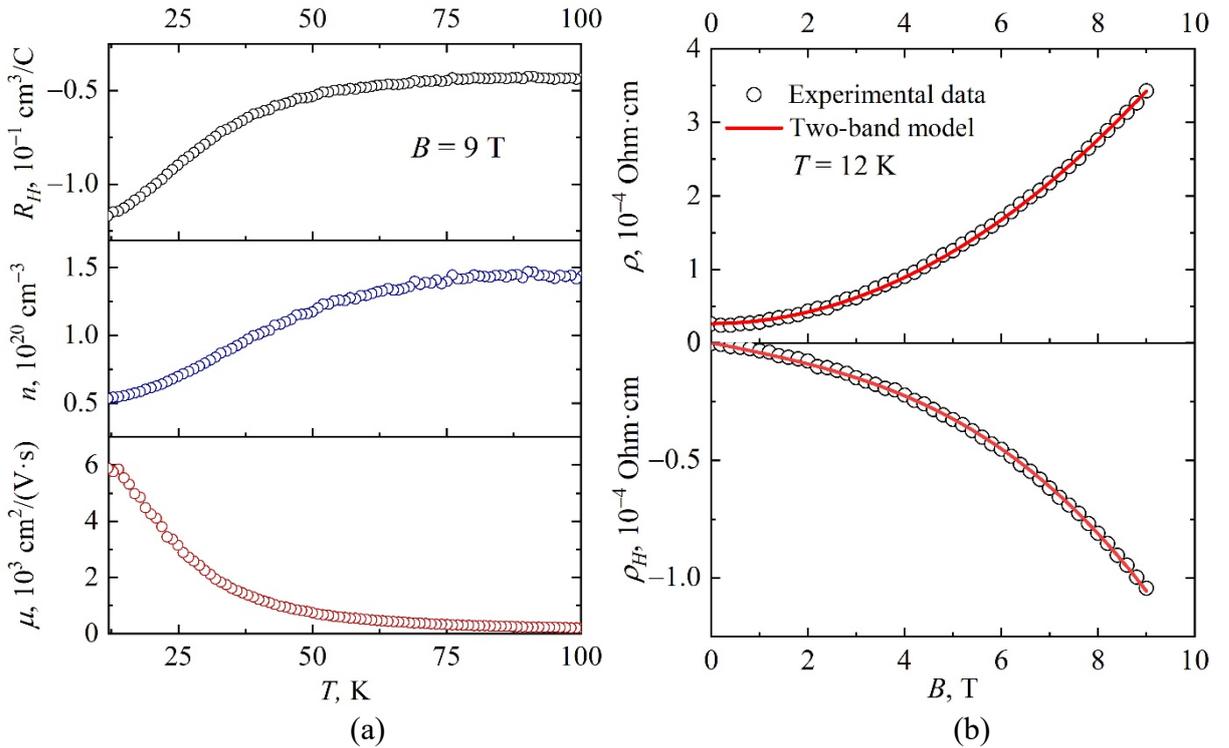

**Figure 7.** (a) The temperature dependences of the Hall coefficient $R_H$, concentration $n$, and mobility $\mu$ of current carriers in WTe$_2$ according to a single-band model in a magnetic field of $B = 9$ T. (b) Field dependences of resistivity $\rho(B)$ in the magnetic field and Hall resistivity $\rho_H(B)$ for WTe$_2$ at $T = 12$ K: open circles show the experimental data; the solid red lines are curves obtained within the framework of the two-band model.